# A Neutrino-Factory Muon Storage Ring to Provide Beams for Multiple Detectors Around the World[*]

D.B. Cline, Y. Fukui, and A. Garren

*Center for Advanced Accelerators*
*Department of Physics and Astronomy, Box 951547*
*University of California, Los Angeles, CA 90095-1547 USA*

**Abstract.** We briefly discuss the physics motivation for a neutrino factory with varying baseline distances of about 1000 to 9000 km. We describe the amount of non planarity of the storage ring required to service three or four detectors at once. A novel bowtie storage ring is described that could in part provide these beams; a preliminary lattice design is given. We give the space angles between the various detector locations and possible sites for neutrino factories. Finally we describe detectors at the Gran Sasso Laboratory and at a new laboratory near Carlsbad, NM to observe the neutrino interactions with wrong sign leptons.

## INTRODUCTION

The physics potential of a muon storage-ring neutrino factory is large[1-3] and includes
(A) Observation of $\nu_e \to \nu_\mu$ at levels of $10^{-3}$; possible detection of $\nu_e \to \nu_\mu$ and $\nu_e \to \nu_\tau$;
(B) Study of $\nu_\mu \to \nu_\tau$ and $\bar\nu_\mu \to \bar\nu_\tau$;
(C) Possible CP violation test for
   (i) $\nu_\mu \to \nu_\tau$, $(\bar\nu_\mu \to \bar\nu_\tau)$;
   (ii) $\nu_e \to \nu_\tau$, $(\bar\nu_e \to \bar\nu_\tau)$, $\nu_e \to \nu_\mu$, $(\bar\nu_e \to \bar\nu_\mu)$;
(D) Possible study of the combined effects $\nu_\mu, \bar\nu_\mu$ to $\nu_e, \bar\nu_e$ final states (T violation).

In the case (Ci), we expect a very small signal in the three-neutrino "CKM" matrix formulation. However, if there are sterile neutrinos, the CP violation could still be very large in this channel.

One strength of the ICANOE[4] detector is the real-time identification of

$$\nu_\tau + N \to \tau + x$$
$$\quad\quad\quad\quad\quad\hookrightarrow e,\mu$$
$$\quad\quad\quad\quad\quad\hookrightarrow \text{hadron}.$$

Using NOMAD-like cuts in addition, the process $\nu_e \to \nu_\mu$, $\nu_\mu + N \to \mu^- + x$ can be detected using a magnetized toroid system such as is employed for ICANOE. We also describe a magnetized ion detector for a new underground laboratory at Carlsbad, NM.

---

In order to address the detection of the processes described before and the resulting background, we need to specify the neutrino factory parameters.

We point out for historical reference that one of the earliest papers on a $\mu^{\pm}$ neutrino source was by Cline and Neuffer in 1979-80.[1] The recent interest in this subject is due to the probable discovery of neutrino oscillations by Superkamiokande (SK) using atmospheric neutrinos.

The current situation in neutrino mass and mixing physics lends itself to some confusion. We do not know if there are sterile neutrinos or not, and a neutrino factory could be the key to resolving this issue.

Another key possibly is to operate a neutrino factory below $\nu_e \to \nu_\tau$ threshold and then observe $\bar{\nu}_e \to \bar{\nu}_\mu \to \mu^-$ in the detector. The detection of $\nu_e \to \nu_\tau$ or $\nu_e \to \nu_\mu$ will be crucial for the use of a neutrino factory to study the neutrino CKM matrix and to search for CP or T violation.

## Backgrounds to Neutrino Oscillation Measurements

The basic concept of the neutrino factory is to detect an oscillation in the non-$\nu_\mu$ channel. For example for a $\mu^-$ storage ring, $\mu^- \to e^- + \nu_\mu + \bar{\nu}_e$. So we have $\bar{\nu}_e$ in the $\nu_\mu$ beam, oscillation into $\bar{\nu}_e \to \bar{\nu}_\mu$, followed by $\bar{\nu}_\mu \to \mu^+$ given a "wrong sign lepton." In the case of $\nu_e \to \nu_\tau$, as we will show, the produced $\tau$ must decay into a $\mu^{\pm}$ channel to measure the lepton charge.

We now discuss the possible level of the signals for neutrino oscillations at a neutrino factory. We can write the transition probably as

$$P_{\mu\tau} = \cos^2\theta_{13} \sin^2 2\theta_{23} \sin^2\left[1.27\left(\frac{L}{E}\right)\Delta^2 m\right] ,$$

$$P_{e\tau} = \sin^2\theta_{23} \sin^2 2\theta_{13} \sin^2\left[1.27\left(\frac{L}{E}\right)\Delta^2 m\right] ,$$

where L is in km, E is in GeV, and $\Delta^2 m$ is in eV$^2$. The SuperKamiokande results suggest that $\theta_{23} \approx 45°$. The current limits on $\nu_\mu \to \nu_e$ constrain $\theta_{13}$ to be less than 20°. Since we know nothing about the neutrino CKM matrix, $\theta_{13}$ could be very small. We can express the $\nu_e \to \nu_\tau$ transition as $P_{e\tau} \approx P_{e\mu} (\cos^2\theta_{23}/\sin^2\theta_{23})$. For $\theta_{23} \approx 45°$, we find $P_{e\tau} \approx P_{e\mu}$). Thus for the issue of background estimates, if $\theta_{13}$ is small, then both $P_{e\tau}$ and $P_{e\mu}$ will be small.

There are natural sources of wrong sign leptons produced in high-energy neutrino interactions. While these backgrounds will not likely affect the detection of $\nu_e \to \nu_\mu$, we will show that for some parameters this background will be dangerous for $\nu_e \to \nu_\tau$ detection.

To show why wrong-sign leptons produced in neutrino interactions may be a limiting background for $\nu_e \to \nu_\tau$, we consider the following:

$$\nu_e \to \nu_\tau \quad , \quad (P_{e\tau}) \; ;$$

$$\nu_\tau + N \to \tau^- + X \quad , \quad (\sigma_\tau/\sigma_{all}) \; ;$$

$$\tau \to \mu^- + X \quad , \quad (P \sim 10^{-1}) \quad .$$

(Note that the $\mu^-$ energy spectrum will be soft.)

The $\mu^\pm$ must be detected to identify the "wrong sign $\tau$" from the large rate of production from $\bar{\nu}_\mu \to \bar{\nu}_\tau$ in the neutrino factory beam. For the parameters $P_{e\tau} \sim 5 \times 10^{-3}$, $\sigma_\tau/\sigma_{all} \sim 5 \times 10^{-2}$. We find the overall probability to get a wrong sign $\mu^\pm$ from an ($e\tau$) oscillation to be $2.5 \times 10^{-5}$.[5,6] We will show that this is below some of the "natural" wrong sign production in neutrino interactions.

The production of charm can lead to wrong sign leptons through the process

$$\nu_\mu + N \to \mu^- + \text{charm}$$
$$\qquad\qquad \hookrightarrow \mu^+ + \nu + X \quad ,$$

with the $\mu^-$ being missed. (This process was first observed by the HPWF group at FNAL in 1974.) We do not expect this to be an important background for $\nu_e \to \nu_\mu$ because of the high energy of the resulting $\mu^-$. However, for $\nu_e \to \nu_\tau$, the $\tau$ decay into a $\mu^-$ channel will result in a soft $\mu^-$, as well as some missing neutrino energy and the same hadronic structure as charm production.

## A BOWTIE MUON RING LATTICE FOR NEUTRINO PRODUCTION

This paper also concerns the possible use of a single storage ring to illuminate three or more detectors around the world, as shown in Fig. 1.

A bowtie-shaped muon storage-ring lattice is discussed. This planar ring can be designed and oriented to send neutrino beams to any two detector sites. Bypasses could be added to send reduced intensity beams to one or two additional sites. There are several advantages of the bowtie shape:

- No net bending, so the polarization is preserved,

- With the ring in a tilted plane, both long straight sections point down into the earth, so neutrinos are sent into two distant detectors. Triangular rings share this advantage.

There are also some disadvantages of the bowtie shape:

- Extra bending required, by 4/3 in this example,

- Geometry constrains the ratio of short and long straight sections, which may increase circumference,

- Complications due to crossing beamlines.

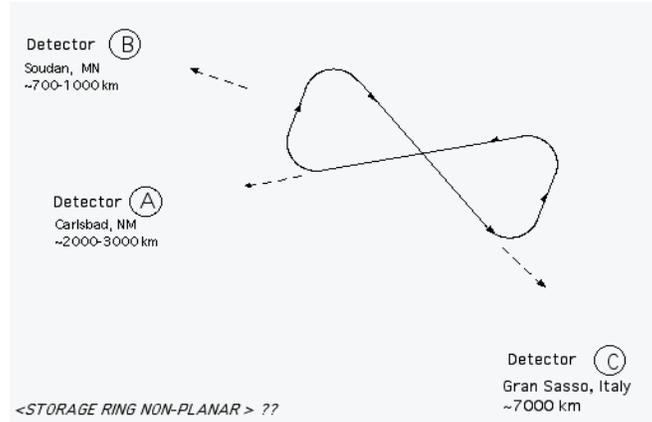

**FIGURE 1.** A scheme to provide neutrino-factory beams to several detectors around the world.

## Lattice Description

The lattice contains three parts: a short straight section, an arc, and a long straight section. The description follows one quarter of the ring, starting at the center of the short straight section on the left side of the figure, and ends at the crossing point at the center of the bowtie (Fig. 2).

### *Short Straight Section*

Half of the short straight section consists of two 14-m arc cells without dipoles. It can also be configured to provide a 20-m free space for injection. One short straight section may be used for injection, the other for RF.

### *Arc*

The arc contains eight FODO cells, two without dipoles. The cell phase advances are 60 deg, and the dipole-free cells act as dispersion suppressors. Twelve 5-m long dipoles each bend the beam 10 deg, so the arc has 120 deg of bending. This angle is needed to cross the beamlines by 60 deg (a typical angle, whose exact value depends on the selection of ring and detector sites).

### *Long Straight Section*

The long dispersion-free straight section provides a muon beam whose decaying muons generate low divergence neutrinos. Two configurations are shown in Figs. 2 and 3. In Fig. 2, the long straight section has quadrupoles in the center (around the crossing point) making two beam waists, each with 50-m beta function values. In Fig. 3, a 200-m magnet-free beamline is provided, with a beam waist at the center with 100-m beta values.

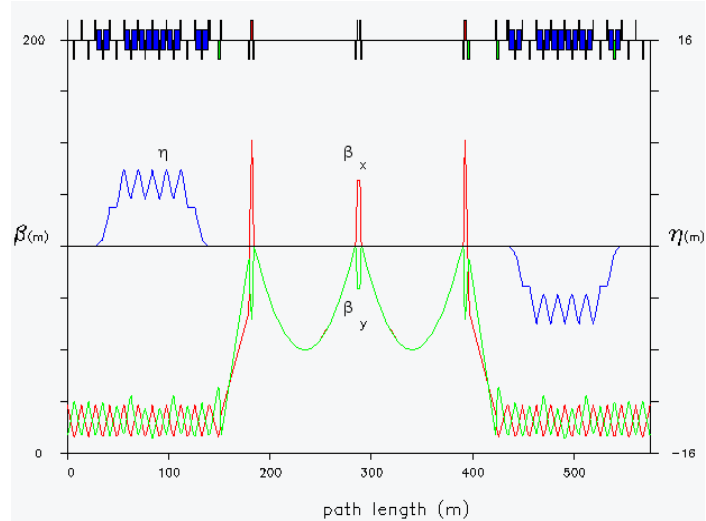

**FIGURE 2.** Half of a bowtie ring lattice. The center of the figure is a straight section with quadrupoles at its center, making two beam waists.

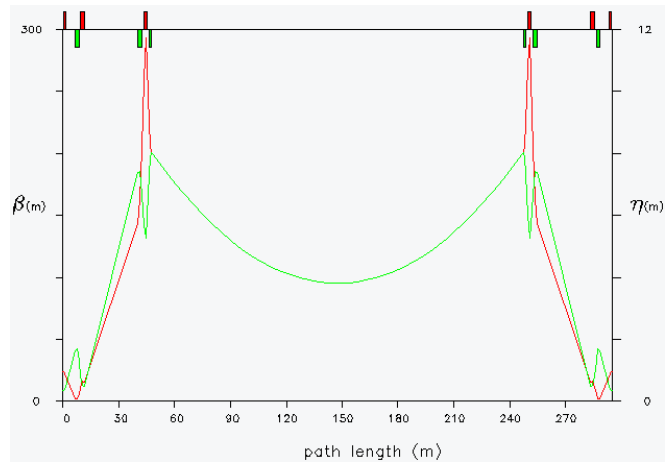

**Figure 3.** Second configuration for long-straight-section insertion of bowtie ring. There are no quadrupoles around the center, where the beams cross.

# Bypasses

A racetrack ring can be configured to deliver one neutrino beam to an arbitrary detector site. A bowtie ring can be configured to deliver neutrino beams from an arbitrary ring site to two arbitrarily chosen detector sites. This can be done by appropriate choice of the ring plane, the orientation of the ring in that plane, and the angle at the crossing point between the two long straight sections (Fig. 4).

Additional detector sites might be accessible from a single muon storage-ring source by inclusion of bypasses. A bypass would begin and end on one of the long straight sections, but lie in a plane that includes the original long straight section but differs from that of the ring. Its magnets would be powered when one desires to send the muons along the deformed bypass path rather than along the normal straight path.

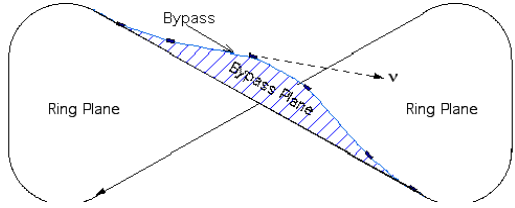

**FIGURE 4.** A possible bypass system to provide neutrino beams to off-angle detectors.

In such a bypass, dipoles would produce a roughly triangular path in the bypass plane, one of whose sides would point to the desired detector. The two necessary degrees of freedom are provided by the angle between the bypass and ring planes and by the magnitude of the deflection given by the bypass dipoles.

Pairs of dipoles should be placed in FODO cells, 180 deg apart, in order to suppress the dispersion.

## SPACE ANGLES FOR THE VARIOUS DETECTOR LOCATIONS AROUND THE WORLD

The ν oscillation probability, $P$, is given by $P = \sin^2(2\theta) \sin^2(1.27 \Delta m^2 L/E_\nu)$, where $\theta$ is the mixing angle, and $\Delta m^2$ $(eV^2)/c^4$ is the difference of mass squared, $L$ km is the baseline length, and $E_\nu$ (GeV) is the ν energy, in a two-flavor ν oscillation model. Assuming reasonably intense ν flux at the far detectors, we obtain wider window ranges of [$\sin^2(2\theta) \sin^2$, $\Delta m^2$] by shooting multiple far ν detectors with mixed baseline lengths.

We discuss space angles which are associated with multiple baselines, and ν flux as a function of the polar angle from the straight section axis of a bow-tie storage ring.

### Baseline

Because of the Earth's rotation, the equator radius is longer than the polar radius by 21 km, which corresponds to 0.3% of the average Earth radius. Figure 5 shows the Earth surface position as functions of the distance from the equator plane and from the rotation axis. The change of the equator radius due to the gravity of the moon and the sun are, respectively, 0.36 m and 0.17 m, depending on the position relationship of the moon, the sun, and the Earth.

Table 1 shows baseline length, vertical angle, and the angle to the North direction at the storage ring site and at the detector site. The numbers in square brackets are the differences between a value with an elliptical radius and a value with a spherical radius.

### Space Angles

Figure 5 shows the baselines and vertical angles, and angles to the North direction at storage rings at Fermilab.

**Table. 1.** Baseline Lengths and Surface Angles.

| | Baseline Length | Vertical Angle* (deg) | Angle* to North (deg) |
|---|---|---|---|
| Fermilab - Soudan, MN | 731.04 [-0.39] km | | |
|   at Fermilab | | 3.473 [ 0.182]** | -23.625 [ 0.073] |
|   at Soudan | | 3.126 [-0.164] | 153.597 [ 0.073] |
| Fermilab - Carlsbad, NM | 1748.87 [-0.18] km | | |
|   at Fermilab | | 7.791 [-0.098] | -121.626 [-0.081] |
|   at Carlsbad | | 8.009 [ 0.120] | 48.689 [-0.075] |
| Fermilab - Gran Sasso, It | 7332.89 [-2.82] km | | |
|   at Fermilab | | 35.149 [ 0.008] | 50.087 [ 0.004] |
|   at Gran Sasso | | 35.155 [ 0.005] | -50.749 [-0.006] |
| Fermilab - Kamioka, Jpn | 9133.63 [-1.87] km | | |
|   at Fermilab | | 45.811 [ 0.008] | -35.178 [-0.016] |
|   at Kamioka | | 45.828 [ 0.026] | 32.102 [ 0.007] |
| BNL - Soudan, MN | 1715.25 [-0.87] km | | |
|   at BNL | | 7.826 [ 0.093] | -56.330 [ 0.076] |
|   at Soudan | | 7.665 [-0.081] | 110.081 [ 0.076] |
| BNL - Carlsbad, NM | 2903.08 [-0.20] km | | |
|   at BNL | | 13.118 [-0.048] | -98.421 [-0.053] |
|   at Carlsbad | | 13.239 [ 0.065] | 62.563 [-0.041] |
| BNL - Gran Sasso, It | 6526.73 [-2.30] km | | |
|   at BNL | | 30.825 [ 0.011] | 56.812 [ 0.002] |
|   at Gran Sasso | | 30.830 [ 0.001] | -59.240 [-0.010] |
| BNL - Kamioka, Jpn | 9635.70 [-1.66] km | | |
|   at BNL | | 49.157 [ 0.014] | -24.150 [-0.012] |
|   at Kamioka | | 49.169 [ 0.027] | 22.571 [ 0.008] |
| CERN - Soudan, MN | 6568.46 [-4.40] km | | |
|   at CERN | | 31.053 [ 0.000] | -48.743 [ 0.007] |
|   at Soudan | | 31.043 [-0.009] | 50.750 [ 0.000] |
| CERN - Carlsbad, NM | 8135.78 [-1.88] km | | |
|   at CERN | | 39.677 [-0.013] | -53.679 [-0.028] |
|   at Carlsbad | | 39.730 [ 0.041] | 41.287 [-0.008] |
| CERN - Gran Sasso, It | 727.58 [-0.37] km | | |
|   at CERN | | 3.120 [-0.107] | 122.679 [ 0.089] |
|   at Gran Sasso | | 3.437 [ 0.114] | -52.084 [ 0.089] |
| CERN - Kamioka, Jpn | 8737.07 [-2.91] km | | |
|   at CERN | | 43.301 [-0.007] | 37.574 [ 0.016] |
|   at Kamioka | | 43.333 [ 0.029] | -31.479 [ 0.003] |

  *positive = North/East; negative = West  
  **[elliptic Earth - spherical Earth]

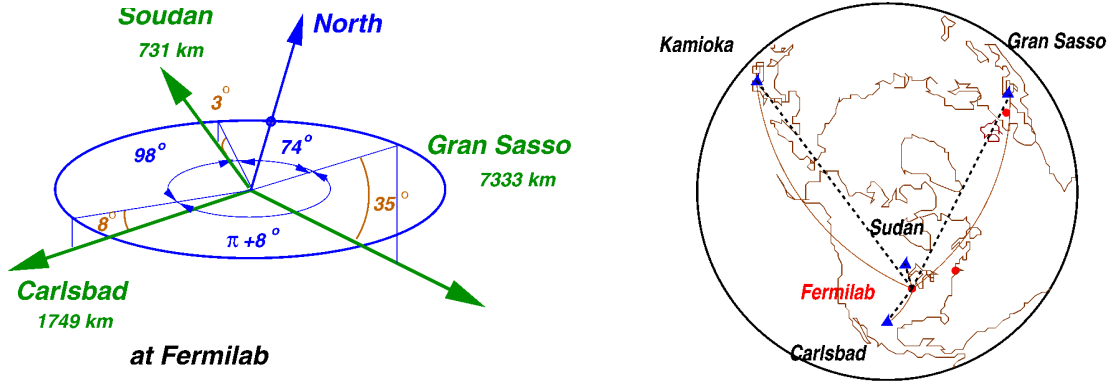

**FIGURE 5.** Baselines and surface angles at Fermilab.

## DETECTORS AT THE GRAN SASSO AND CARLSBAD SITES

We assume that distances of 2000 to 7000 km are useful to study neutrino oscillations, matter effects, and possibly CP violation. We also assume that the detectors must be located underground, because of the enormous backgrounds from cosmic rays. Finally, we assume that existing underground facilities will be used for the detector. In this section, we discuss detector locations of ~ 2000 to 3000 km from either FNAL or BNL and 6000 to 7000 km to the LNGS.

We discuss two detectors:

(1) ICANOE at the LNGS [Fig. 6(A) and (B)] and

(2) A magnetized Fe detector at the Carlsbad Underground Laboratory (Fig. 7).

These two detectors are in a sense the extremes of the detectors one may use for a neutrino factory. The ICANOE detector will have excellent electron identification and can observe $\nu_{\mu,\tau} \to \nu_e$, whereas the Carlsbad detector will likely be used for wrong-sign muon identification, i.e., $\bar{\nu}_e \to \bar{\nu}_\mu \to \mu^+$ in the detector. To search for CP violation, both detectors will have to use $\mu^\pm$ sign determination, since we have to way to measure the electric charge of the "$e$"-like signals.

We will first address the issue of the crossing angle of the neutrino beam on the detector. This will not be a problem for a new hall as can be constructed at the Carlsbad site, but as shown in Fig. 6(B), it could be a problem for a fixed laboratory like the LNGS. It is possible that part of the ICANOE detector can be reconfigured to fit this space angle.

## SUMMARY

We considered the option of shooting multiple far detectors with intense $\nu$ beam from a bow-tie storage ring. The base line lengths, surface angles, and space angles were calculated with both elliptical and spherical Earth models. The transverse angle of the muons in the straight section of a bow-tie storage ring is demonstrated to be much smaller than the HWHM (half width half maximum) of the $\nu$ angle from the axis. We also discussed the important neutrino physics that can be carried out with detectors at LNGS (ICANOE) and at Carlsbad, NM.

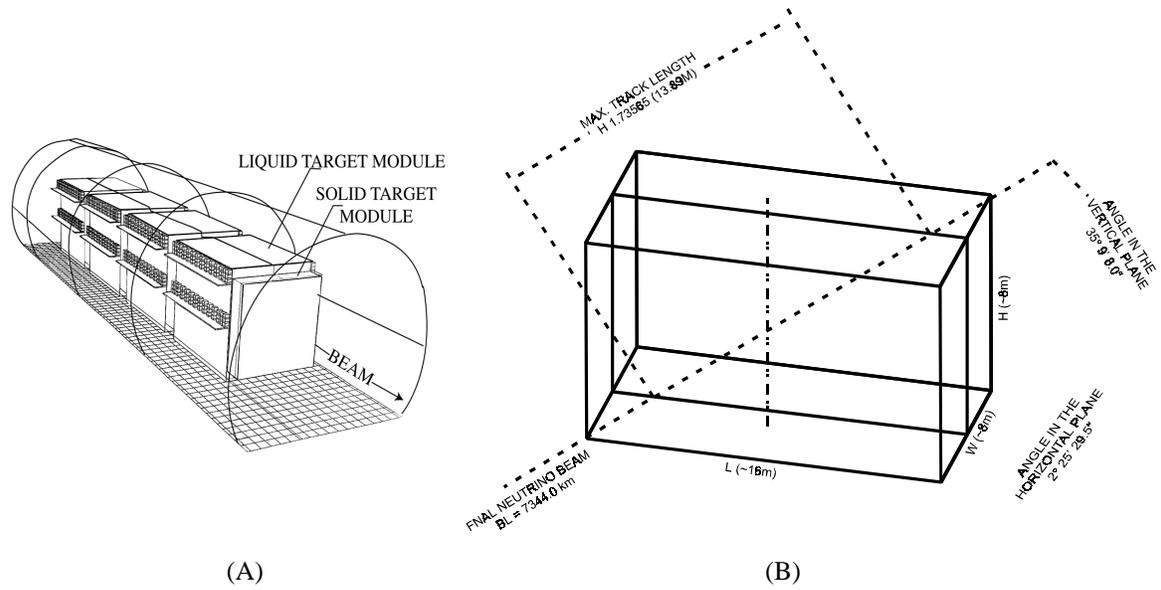

**FIGURE 6.** (A) Schematic of the ICANOE detector to be constructed at the LNGS; (B) Detector crossing of the LBL beam from FNAL for the ICANOE detector at LNGS.

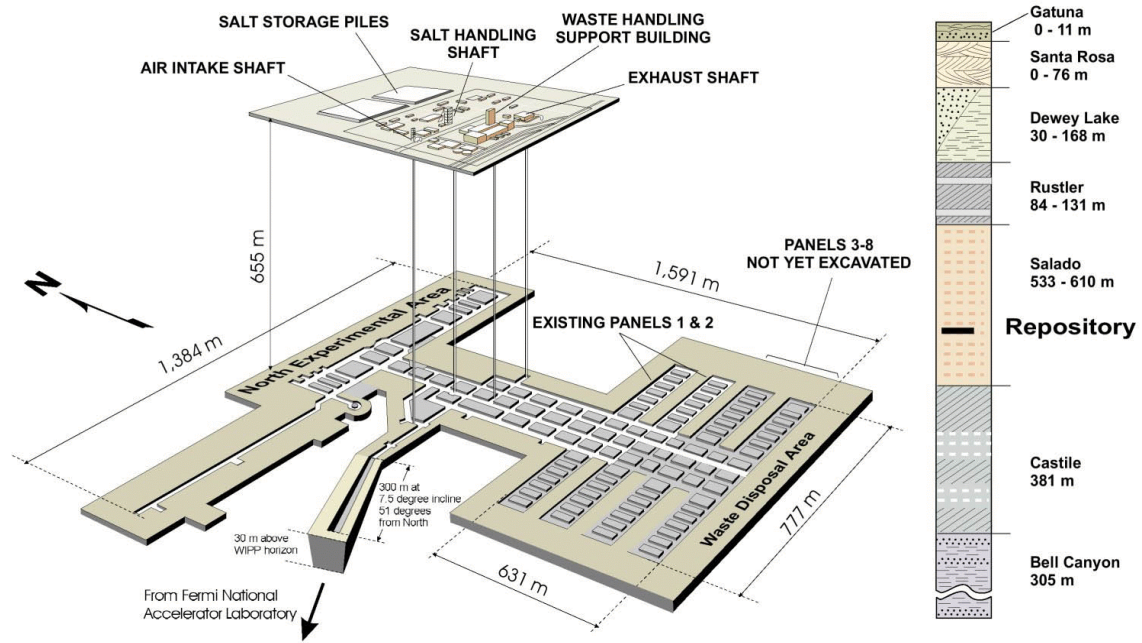

**FIGURE 7.** Possible neutrino-factory detector at the WIPP site in Carlsbad, NM.

# ACKNOWLEDGEMENTS

We wish to thank J. Gallardo, R. Nelson, R. Palmer, and F. Sergiampietri for their help and encouragement.